\documentclass[submission,copyright,creativecommons]{eptcs}
\providecommand{\event}{Postproceedings of ITRS'14} 
\usepackage{breakurl}             
\usepackage{amsmath}
\usepackage{listings}
\usepackage{todonotes}
\usepackage{graphicx}
\usepackage{caption}
\usepackage{subcaption}
\usepackage{color}
\newdimen\proofrulebreadth \proofrulebreadth=.05em
\newdimen\proofdotseparation \proofdotseparation=1.25ex
\newdimen\proofrulebaseline \proofrulebaseline=2ex
\newcount\proofdotnumber \proofdotnumber=3
\let\then\relax
\def\hfi{\hskip0pt plus.0001fil}
\mathchardef\squigto="3A3B
\newif\ifinsideprooftree\insideprooftreefalse
\newif\ifonleftofproofrule\onleftofproofrulefalse
\newif\ifproofdots\proofdotsfalse
\newif\ifdoubleproof\doubleprooffalse
\let\wereinproofbit\relax
\newdimen\shortenproofleft
\newdimen\shortenproofright
\newdimen\proofbelowshift
\newbox\proofabove
\newbox\proofbelow
\newbox\proofrulename
\def\shiftproofbelow{\let\next\relax\afterassignment\setshiftproofbelow\dimen0 }
\def\shiftproofbelowneg{\def\next{\multiply\dimen0 by-1 }%
\afterassignment\setshiftproofbelow\dimen0 }
\def\setshiftproofbelow{\next\proofbelowshift=\dimen0 }
\def\setproofrulebreadth{\proofrulebreadth}

\def\prooftree{
\ifnum  \lastpenalty=1
\then   \unpenalty
\else   \onleftofproofrulefalse
\fi
\ifonleftofproofrule
\else   \ifinsideprooftree
        \then   \hskip.5em plus1fil
        \fi
\fi
\bgroup
\setbox\proofbelow=\hbox{}\setbox\proofrulename=\hbox{}%
\let\justifies\proofover\let\leadsto\proofoverdots\let\Justifies\proofoverdbl
\let\using\proofusing\let\[\prooftree
\ifinsideprooftree\let\]\endprooftree\fi
\proofdotsfalse\doubleprooffalse
\let\thickness\setproofrulebreadth
\let\shiftright\shiftproofbelow \let\shift\shiftproofbelow
\let\shiftleft\shiftproofbelowneg
\let\ifwasinsideprooftree\ifinsideprooftree
\insideprooftreetrue
\setbox\proofabove=\hbox\bgroup$\displaystyle 
\let\wereinproofbit\prooftree
\shortenproofleft=0pt \shortenproofright=0pt \proofbelowshift=0pt
\onleftofproofruletrue\penalty1
}

\def\eproofbit{
\ifx    \wereinproofbit\prooftree
\then   \ifcase \lastpenalty
        \then   \shortenproofright=0pt  
        \or     \unpenalty\hfil         
        \or     \unpenalty\unskip       
        \else   \shortenproofright=0pt  
        \fi
\fi
\global\dimen0=\shortenproofleft
\global\dimen1=\shortenproofright
\global\dimen2=\proofrulebreadth
\global\dimen3=\proofbelowshift
\global\dimen4=\proofdotseparation
\global\count255=\proofdotnumber
$\egroup  
\shortenproofleft=\dimen0
\shortenproofright=\dimen1
\proofrulebreadth=\dimen2
\proofbelowshift=\dimen3
\proofdotseparation=\dimen4
\proofdotnumber=\count255
}

\def\proofover{
\eproofbit 
\setbox\proofbelow=\hbox\bgroup 
\let\wereinproofbit\proofover
$\displaystyle
}%
\def\proofoverdbl{
\eproofbit 
\doubleprooftrue
\setbox\proofbelow=\hbox\bgroup 
\let\wereinproofbit\proofoverdbl
$\displaystyle
}%
\def\proofoverdots{
\eproofbit 
\proofdotstrue
\setbox\proofbelow=\hbox\bgroup 
\let\wereinproofbit\proofoverdots
$\displaystyle
}%
\def\proofusing{
\eproofbit 
\setbox\proofrulename=\hbox\bgroup 
\let\wereinproofbit\proofusing
\kern0.3em$
}

\def\endprooftree{
\eproofbit 
  \dimen5 =0pt
\dimen0=\wd\proofabove \advance\dimen0-\shortenproofleft
\advance\dimen0-\shortenproofright
\dimen1=.5\dimen0 \advance\dimen1-.5\wd\proofbelow
\dimen4=\dimen1
\advance\dimen1\proofbelowshift \advance\dimen4-\proofbelowshift
\ifdim  \dimen1<0pt
\then   \advance\shortenproofleft\dimen1
        \advance\dimen0-\dimen1
        \dimen1=0pt
        \ifdim  \shortenproofleft<0pt
        \then   \setbox\proofabove=\hbox{%
                        \kern-\shortenproofleft\unhbox\proofabove}%
                \shortenproofleft=0pt
        \fi
\fi
\ifdim  \dimen4<0pt
\then   \advance\shortenproofright\dimen4
        \advance\dimen0-\dimen4
        \dimen4=0pt
\fi
\ifdim  \shortenproofright<\wd\proofrulename
\then   \shortenproofright=\wd\proofrulename
\fi
\dimen2=\shortenproofleft \advance\dimen2 by\dimen1
\dimen3=\shortenproofright\advance\dimen3 by\dimen4
\ifproofdots
\then
        \dimen6=\shortenproofleft \advance\dimen6 .5\dimen0
        \setbox1=\vbox to\proofdotseparation{\vss\hbox{$\cdot$}\vss}%
        \setbox0=\hbox{%
                \advance\dimen6-.5\wd1
                \kern\dimen6
                $\vcenter to\proofdotnumber\proofdotseparation
                        {\leaders\box1\vfill}$%
                \unhbox\proofrulename}%
\else   \dimen6=\fontdimen22\the\textfont2 
        \dimen7=\dimen6
        \advance\dimen6by.5\proofrulebreadth
        \advance\dimen7by-.5\proofrulebreadth
        \setbox0=\hbox{%
                \kern\shortenproofleft
                \ifdoubleproof
                \then   \hbox to\dimen0{%
                        $\mathsurround0pt\mathord=\mkern-6mu%
                        \cleaders\hbox{$\mkern-2mu=\mkern-2mu$}\hfill
                        \mkern-6mu\mathord=$}%
                \else   \vrule height\dimen6 depth-\dimen7 width\dimen0
                \fi
                \unhbox\proofrulename}%
        \ht0=\dimen6 \dp0=-\dimen7
\fi
\let\doll\relax
\ifwasinsideprooftree
\then   \let\VBOX\vbox
\else   \ifmmode\else$\let\doll=$\fi
        \let\VBOX\vcenter
\fi
\VBOX   {\baselineskip\proofrulebaseline \lineskip.2ex
        \expandafter\lineskiplimit\ifproofdots0ex\else-0.6ex\fi
        \hbox   spread\dimen5   {\hfi\unhbox\proofabove\hfi}%
        \hbox{\box0}%
        \hbox   {\kern\dimen2 \box\proofbelow}}\doll%
\global\dimen2=\dimen2
\global\dimen3=\dimen3
\egroup 
\ifonleftofproofrule
\then   \shortenproofleft=\dimen2
\fi
\shortenproofright=\dimen3
\onleftofproofrulefalse
\ifinsideprooftree
\then   \hskip.5em plus 1fil \penalty2
\fi
}

\input{Macros}

\title{Typing Classes and Mixins with Intersection Types}
\author{
Jan Bessai\qquad Boris D\"udder \qquad Andrej Dudenhefner
\institute{Technical University of Dortmund, Germany}
\email{\{jan.bessai, boris.duedder, Andrej.dudenhefner\}@cs.tu-dortmund.de}
\and
Tzu-Chun Chen
\institute{Technical University of Darmstadt, Germany}
\email{tcchen@rbg.informatik.tu-darmstadt.de}
\and
Ugo de'Liguoro
\institute{University of Torino, Italy}
\email{ugo.deliguoro@unito.it}
}
\def\titlerunning{Classes and Mixins with Intersection Types}
\def\authorrunning{J.Bessai,A.Dudenhefner,B.D\"{u}dder,T.Chen,U.de'Liguoro}
\begin{document}
\maketitle

\begin{abstract}
We study an assignment system of intersection types for a lambda-calculus with records and a record-merge operator, where types
are preserved both under subject reduction and expansion. The calculus
is expressive enough to naturally represent mixins as functions over recursively defined classes, whose fixed points,  the objects, 
are recursive records. In spite of the double recursion that is involved in their definition, classes and mixins can be meaningfully typed
without resorting to neither recursive nor F-bounded polymorphic types. 

We then adapt mixin construct and composition to Java and C\#, relying solely on existing features in such a way that the resulting code remains typable in the respective type systems. We exhibit some example code, and study its typings in the intersection type system via interpretation into the lambda-calculus with records we have proposed.
\end{abstract}

\section{Introduction}

Starting with Cardelli's pioneering work \cite{Cardelli84}, various typed lambda calculi extended with records have been thoroughly studied to model  sophisticated features of object-oriented programming languages, like recursive objects and classes, object extension, method overriding and inheritance (see e.g. \cite{AbadiCardelli,BruceBook,KiselyovLaemmel05}).

Here we focus on object composition based on mixins, a study initiated in \cite{BrachaThesis,BrachaC90} and based on the recursive record model and F-bounded polymorphic types \cite{CHOM89,CookHC90}.
In the object-oriented paradigm, mixins have been introduced as an alternative construct w.r.t. class inheritance to avoid semantic ambiguities caused by multiple inheritance. 
Together with abstract classes and traits, mixins can be considered as an advanced construct to obtain flexible implementation of module libraries and to enhance code reusability; many popular programming languages miss native support for mixins, but they are object of intensive study and research (e.g. \cite{BonoPS99,OderskyZ05}). Scala supports mixins, which recently have been used to model feature oriented programming \cite{OliveiraSLC13}. Dynamic languages, such as JavaScript, offer multiple ways to compensate for the lack of mixins \cite{Osmani12}. 

In \cite{JR13} a new technique for synthesizing programs from components has been based on combinatory logic and intersection types.
Aiming at a future extension of the approach to Object-Oriented (OO) libraries, we study an assignment system of intersection types for a $\lambda$-calculus with records and a record-merge operator, building over and extending the systems proposed in \cite{deLiguoro01,Bakeld08,RoweB14}.

In section 2, we propose a type-free $\lambda$-calculus with record and record-merge operator, and define a Curry-style type assignment system. Such a system
is polymorphic in the sense that any term has infinitely many types, including the trivial type $\omega$. In doing that, we look for an extension of the system in \cite{BCD}, and in particular we expect that typing is invariant under subject reduction and expansion. In this respect
the main difficulty consists in the typing of Bracha-Cook's merge operator, which embodies at the same time record extension and field overriding.
To solve such a problem we distinguish among ordinary $\lambda$-terms and record terms, allowing just record terms to be merged to arbitrary terms.
In section 3, we then study the typings in our system of the encoding of classes and mixins into the $\lambda$-calculus proposed in  \cite{CookHC90,BrachaC90},
identifying a subset of intersection types that can be seen as meaningful types for such entities.
We then consider (Section 3) the problem of using the proposed system in the case of the actual programming languages Java and C\#, that do not embody mixins in their syntax; nevertheless programs that typecheck w.r.t. the type system of the respecitive programming language can be seen as the product of the fully applied mixin compositions. We face these challenges by using delegation instead of inheritance. This gives rise to a design pattern that can be interpreted in the $\lambda$-calculus with records, such that the typings we have devised for classes and mixins in the theoretical model apply to actual programming languages.

This work evolves from the contributions \cite{BessaiDDM14,UdLTC2014} to the workshop ITRS'14. Joining the contributions from both previous papers, this work contributes a way to model classes and mixins typable in the intersection type system. The flexibility of this type system allows to augment types by semantic features, required for a future extension to automatically synthesize OO code \cite{BessaiDDM14}. Furthermore we present a correspondence between the model and actual programming languages.

\noindent\emph{The authors have been supported by the research grant of the European Union ICT COST Action (IC1201): Behavioural Types for Reliable Large-Scale Software Systems (BETTY).}


\newcommand{\const}{\textsf{t}}
\newcommand{\mixin}{M}
\newcommand{\lbl}{\mbox{\it lbl}}

\section{Intersection types for a $\lambda$-calculus with records and merge}
\label{sec:intersection-types-lambda-rec}

We consider a type-free $\lambda$-calculus of records equipped with a  merge operator.
The syntax of $\lambdaR$ is defined by the following grammar:
\[
\begin{array}{c}
 \mbox{Term:} ~~~ M,N     ::=  x \mid (\lambda x.M) \mid (MN) \mid R  \mid (M.a)  \mid (M \oplus R) \qquad\mbox{Record:} ~~~
R   ::=  \record{ a_i = M_i \ | \ i \in I }
\end{array}
\]
where $x\in\Variable$ and $a\in\Label$ range over denumerably many variables and labels respectively.
In $R \equiv \record{ a_i = M_i \ | \ i \in I }$ (writing $\equiv$ for syntactic identity) 
the set $I$ is finite; we define $\lbl(R) = \Set{a_i \mid i\in I}$ to denote the set of labels of $R$
and we use the abbreviation $a=M \in R$ for $a \equiv a_i \And M \equiv M_i$ for some $i\in I$.

\begin{definition}[$\lambdaR$ Reduction]
\[\begin{array}{lrcl}
(\beta) & \ (\lambda x.M)N &\reduces & M\Subst{N}{x} \\ [1mm]
(\redSel) & \Pair{a_i=M_i \ \mid \  i\in I}.a_j & \reduces & M_j  \hspace{4cm} \mbox{if $j \in I$} \\ [1mm]
(\redMergeThree) & \Pair{a_i=M_i \ \mid \ i\in I} \Override \Pair{a_j=N_j \ \mid \ j\in J} & \reduces &
	\Pair{a_i=M_i, \ a_j=N_j \ \mid \  i\in I\setminus J, \  j\in J}
	
\end{array}\]
\end{definition}
The last reduction makes it apparent that $\Override$ is record merge from \cite{BrachaThesis}, which is written $\leftarrow_r$, and the analogous operator in \cite{BrachaC90},
although in our notation $M\Override R$ the record arguments are listed in the opposite order than in \cite{BrachaC90}, namely whenever $M$ reduces to
a record $R'$ the fields in $R$
prevail over those in $R'$ having the same labels.

\noindent

\begin{definition}[Intersection types for $\lambdaR$]  \label{def:intertype:lambdaR}
\[\sigma,\tau ::= \alpha \mid \const \mid \omega \mid \sigma\arrow\tau \mid \sigma \inter \tau \mid \Pair{a:\sigma}. \]
\end{definition}
Here $\alpha$ ranges over a countable set of type variables, $\const$  are type constants for ground types. e.g. $\Int, \Bool$, or
for (atomic) abstract properties, like ${\sf EvenInt}$ or ${\sf CelsiusDeg}$; $\omega$ is the universal type; 
$\sigma\arrow\tau$ and $\sigma \inter \tau$
are the arrow and the intersection types respectively. 
Finally $\Pair{a:\sigma}$ is the type of records having a label $a$ with value of type $\sigma$.

\begin{definition}[Type Inclusion]\label{def:typeInclusion}
{\em
Over types we consider the preorder $\leq$ extending EATS axioms (see e.g. \cite{AmadioCurien98}, Def. 3.3.1) by the last two axioms:
\begin{align*}
& \sigma \leq \omega, \qquad   \omega \leq \omega\to\omega, \qquad  \sigma\inter\tau \leq \sigma, \qquad && \sigma\inter\tau \leq \tau, \qquad (\sigma\to\tau_1) \inter (\sigma\to\tau_2) \leq \sigma \to \tau_1\inter\tau_2, \\ 
& \sigma \leq \tau_1 \And \sigma\leq \tau_2 \Then \sigma \leq \tau_1\inter\tau_2, &&
\sigma_2 \leq \sigma_1 \And \tau_1 \leq \tau_2 \Then \sigma_1\to\tau_1 \leq \sigma_2\to\tau_2,\\
& \sigma \leq \tau \Then \Pair{a:\sigma} \leq \Pair{a:\tau}, 
 && \Pair{a:\sigma}\inter\Pair{a:\tau} \leq \Pair{a:\sigma\inter\tau}.
\end{align*}
}
\end{definition}

\noindent Note that we have
$\Pair{a:\sigma}\inter\Pair{a:\tau} = \Pair{a:\sigma\inter\tau}$
where $=$ is the symmetric closure of $\leq$. Also we have
$\Pair{a:\omega} \neq \omega$ for any $a$. 

\begin{definition}[Type Assignment]\label{def:typeAssignment}
{\em 
The assignment system adds the rules $(\selRule), (\recRule), (\oplus_l), (\oplus_r)$ to the intersection type assignment \cite{BCD} (but with the $\leq$ relation from Def. \ref{def:typeInclusion}):
\begin{align*}
&\prooftree
	x:\sigma \in \Gamma
	\justifies
	\Gamma \deduces x:\sigma
	\using (\Axiom)
\endprooftree 
\qquad
&&\prooftree
	\Gamma, x:\sigma \deduces M: \tau
	\justifies
	\Gamma \deduces \lambda x.M: \sigma\to\tau
	\using (\ArrI)
\endprooftree 
\qquad
&&\prooftree
	\Gamma \deduces M: \sigma\to\tau \qquad \Gamma \deduces N:\sigma
	\justifies
	\Gamma \deduces MN:\tau
	\using (\ArrE)
\endprooftree \\
~\\
&\prooftree
	\justifies
	\Gamma \deduces M:\omega
	\using (\omega)
\endprooftree 
\qquad 
&&\prooftree
	\Gamma \deduces M: \sigma \quad 
	\Gamma \deduces M: \tau
	\justifies
	\Gamma \deduces M: \sigma\inter\tau
	\using (\IntI)
\endprooftree 
\qquad
&&\prooftree
	\Gamma \deduces M: \sigma \qquad \sigma\leq\tau
	\justifies
	\Gamma \deduces M : \tau
	\using (\leq)
\endprooftree
\end{align*}
\begin{align*}
&\prooftree
	\Gamma \deduces M: \Pair{a:\sigma}
	\justifies
	\Gamma \deduces M.a : \sigma
	\using (\selRule)
\endprooftree 
\qquad
&&\prooftree
	\Gamma \deduces M: \sigma \quad 
	a = M \in R
	\justifies
	\Gamma \deduces R : \Pair{a: \sigma}
	\using (\recRule)
\endprooftree 
\qquad\\
~\\
&\prooftree \Gamma \der M:\Pair{a:\sigma} \qquad a \not\in \lbl(R)
\justifies \Gamma \der M \Override R:\Pair{a:\sigma}
\using (\oplus_l)
\endprooftree 
\qquad
&&\prooftree \Gamma \der R:\Pair{a:\sigma}
\justifies \Gamma \der M \Override R:\Pair{a:\sigma}
\using (\oplus_r)
\endprooftree 
\end{align*}
}\end{definition}
\noindent
Record types express just partial information about the fields in a record, as it is apparent from rule $(\recRule)$;  typings of more than one field in a record are obtained by means of intersection (see below). Rule $(\selRule)$ is the expected one. Rules $(\oplus_l)$ and $(\oplus_r)$ are not symmetric because of the restriction
$a \not\in \lbl(R)$ in the last rule. Such a restriction is essential for the soundnes of the system, in particular for subject reduction to hold. E.g. we have
$
\record{a=M} \Override \record{a=N} \reduces \record{a=N}
$; now supposing that $M:\sigma$, $N:\tau$ where $\sigma\neq\tau$ and that $\sigma$ cannot be assigned to $N$, without the restriction we could type
$\record{a=M} \Override \record{a=N}$ by $\Pair{a:\sigma}$ and even by $\Pair{a:\sigma}\inter\Pair{a:\tau} = \Pair{a:\sigma\inter\tau}$, but
we couldn't type  $\record{a=N}$ by neither of these types.

\medskip
\noindent
Let us abbreviate
$\Pair{a_i:\sigma_i \ \mid \ i\in I} \equiv \bigcap_{i\in I}  \Pair{a_i:\sigma_i}$
where we assume the $a_i$ to be pairwise distinct. By abusing terminology we call it a {\em record type}. Then the following rule is admissible:
\[
\prooftree
	\sigma_j \leq \tau_j \qquad \forall j \in J \subseteq I
	\justifies
	\Pair{ a_i : \sigma_{i} \ \mid \ i \in I } \leq \Pair{a_j : \tau_{j} \ \mid \ j \in J } 
	\endprooftree
\]
which is record subtyping in width and depth. Further we have the following admissible typing rules:
\[
\small
\begin{array}{c}
\prooftree
	\Gamma \deduces N_i:\sigma_i \qquad \forall i \in I \subseteq J
	\justifies
	\Gamma \deduces \Pair{a_i=N_i \ \mid \ i\in J}: \Pair{a_i:\sigma_i \ \mid \  i\in I} 
\endprooftree 
\qquad

\prooftree
	\Gamma \der M : \Pair{a_i:\sigma_i\ \mid \ i \in I} \qquad 
	\Gamma \der R : \Pair{a_j:\tau_j \ \mid \ j \in J}
\justifies
	\Gamma \der M \Override R :  \Pair{a_i:\sigma_i , \ a_j:\tau_j  \ \mid \  i \in I\setminus J, \ j \in J}
\endprooftree
\end{array}
\]

\newpage
 We exemplify the typings of functions from records to records, that are at the earth of the encoding
of mixins. Let $\mixin_{R_i}\equiv \lambda x.(x \Override R_i)$, where
$R_1 \equiv \Pair{a =N_1}$, $R_2 \equiv \Pair{b =N_2}$ and $R_3 \equiv \Pair{a =N_3}$. For simplicity let us
assume $x\not\in\fv(R_i)$ for $i=1,2,3$; then
\[
\prooftree
	\prooftree
		\Gamma, x:\omega \der x:\omega
		\qquad
		\prooftree
			\Gamma, x:\omega \der N_1:\sigma_1
		\justifies
			\Gamma, x:\omega \der \Pair{a =N_1}: \Pair{a:\sigma_1}
		\endprooftree
	\justifies
		\Gamma, x:\omega \der x \Override \Pair{a =N_1}: \Pair{a:\sigma_1}
	\endprooftree
\justifies
	\Gamma \der \mixin_{R_1} \equiv \lambda x.(x \Override \Pair{a =N_1}) : \omega \arrow \Pair{a:\sigma_1}
\endprooftree
\]
and also
\[
\prooftree
	\prooftree
		\Gamma, x:\Pair{b:\sigma_2} \der x: \Pair{b:\sigma_2} 
		\qquad
		b \not\in \lbl(\Pair{a =N_1})
	\justifies
		\Gamma, x:\Pair{b:\sigma_2} \der x \Override \Pair{a =N_1}: \Pair{b:\sigma_2}
	\endprooftree
\justifies
	\Gamma \der \mixin_{R_1} \equiv \lambda x.(x \Override \Pair{a =N_1}) : \Pair{b:\sigma_2} \arrow \Pair{b:\sigma_2}
\endprooftree
\]
therefore by $\omega \arrow \Pair{a:\sigma_1}  \leq \Pair{b:\sigma_2} \arrow \Pair{a:\sigma_1}$
we have
\[
\Gamma \der \mixin_{R_1} : 
	 (\Pair{b:\sigma_2} \arrow \Pair{a:\sigma_1}) \inter (\Pair{b:\sigma_2} \arrow \Pair{b:\sigma_2})
\]
that is
$\Gamma \der \mixin_{R_1} : 
	 \Pair{b:\sigma_2} \arrow  \Pair{a:\sigma_1, b:\sigma_2}$. 
Similarly for $\mixin_{R_2}$, assuming that $\Gamma \der N_2 : \sigma_2$ we have
$\Gamma \der \mixin_{R_2} : \omega\arrow \Pair{b:\sigma_2}$
so that
\[
\prooftree
	\Gamma \der \mixin_{R_2} : \omega\arrow \Pair{b:\sigma_2} \qquad
	\Gamma \der \mixin_{R_1} : \Pair{b:\sigma_2} \arrow  \Pair{a:\sigma_1, b:\sigma_2}
\justifies
	\Gamma \der \mixin_{R_1} \circ \mixin_{R_2}: \omega\arrow \Pair{a:\sigma_1, b:\sigma_2}
\endprooftree
\]
where $M \circ N \equiv \mbox{\bf B}\,M\,N =_\beta \lambda x. M(N\,x)$.
If $\Gamma\der N_1:\sigma_1$, $\Gamma\der N_3:\sigma_3$ but $\Gamma\not\der N_1:\sigma_3$ we have
\[
\Gamma \der \mixin_{R_3} : \omega\arrow \Pair{a:\sigma_3}\
\text{and}\
\Gamma \der \mixin_{R_1} : \omega\arrow \Pair{a:\sigma_1} \leq 
\Pair{a:\sigma_3} \arrow \Pair{a:\sigma_1}
\]
so that
$\Gamma \der \mixin_{R_1} \circ \mixin_{R_3}: \omega\arrow \Pair{a:\sigma_1}$ 
but
$\Gamma \not\der \mixin_{R_1} \circ \mixin_{R_3}: \omega\arrow \Pair{a:\sigma_3}$. 

\mypar
Let $M = N$ be the convertibility relation generated by $\reduces$; then we can prove\footnote{Proof available in the appendix of \url{http://www-seal.cs.tu-dortmund.de/seal/downloads/papers/paper-ITRS2014-postproceedings.pdf}}:
\begin{theorem}[Type Invariance for $\lambdaR$]\label{thr:subjectRedExp} \rm
For any $M, N \in \lambdaR$,
\[\Gamma\deduces M:\sigma ~ \&~M = N ~~ \Then~~ \Gamma\deduces N:\sigma.\]
\end{theorem}

\begin{remark}\label{rem:recordRes} 
{\em By definition $R \Override x \not \in \lambdaR$ and hence
$\lambda x . (R \Override x) \not \in \lambdaR$.
If we admit record terms of the shape $R\Override x$ then their typing would be highly problematic. 
The problem is that  to apply rule $(\oplus_l)$ when typing $R \Override x$ we have to extend $\lbl$ to variables. 
If we set
$\lbl(x) = \emptyset$  the side condition $a\not\in\emptyset$ of rule $(\oplus_l)$ is always satisfied, and 
we could derive for $R\Override x$ all the types of $R$; hence for any $\sigma$:
\[
\prooftree
	\prooftree
		\Gamma, x:\sigma \der R : \Pair{a:\tau}   \qquad a \not \in \lbl(x) = \emptyset
	\justifies
		\Gamma, x:\sigma \der R \Override x: \Pair{a:\tau}
	\endprooftree
\justifies
	\Gamma \der \lambda x. (R \Override x): \sigma \arrow \Pair{a:\tau}
\endprooftree
\]
Let $\sigma \equiv \Pair{a: \rho}$ and $N$ such that
$\Gamma \deduces N : \rho$ and $\Gamma \not \deduces N: \tau$.
Then
$\Gamma\der (\lambda x . (R \Override x)) \Pair{a=N}: \Pair{a:\tau}$, 
while $\Gamma\not\der R\Override \Pair{a=N}:\Pair{a:\tau}$, contradicting subject reduction. 

One could think to change the definition of $\lbl(x)$ to $\lbl'(x)=\Label$, since the variable $x$ can be replaced by $\beta$-reduction with any
record $R'$, whose set of labels we cannot predict. This time the side condition $a\not\in\Label$ never holds true, so that $(\lambda x. (R \Override x) ) \Pair{a=N}$
would have less types than $R\Override \Pair{a=N}$, breaking subject expansion.
%
}\end{remark}
\vspace{-1pt}



\section{Class and mixin interpretation and typings}
\label{sec:interpretations}


%
%

Here we study the typings of the interpretations of classes and mixins based on the recursive-record model, following \cite{BrachaC90}. 
We proceed by steps, considering non-recursive classes and mixins first; then we move to the respective recursive versions.

\mypar {\bf Non-recursive classes.}
Under recursive record interpretations a {\em non-recursive class} is a function:
\[C \defEq \lambda \vec{x} \ \lambda \self. \record{a_1=M_1, \ldots , a_k=M_k}\]
where $\vec{x}$ are the parameters for the initial values (the {\em state}). Class $C$ is pre-instantiated when applied to a tuple of values $\vec{v}$
of the same length as $\vec{x}$:  $C \ \vec{v} =  \lambda \self. \record{a_1=M_1[\vec{v}/\vec{x}], \ldots , a_k=M_k[\vec{v}/\vec{x}]}$.
{\em Objects}, or {\em class instances}, are fixed points of pre-instantiated classes:\\
$O \defEq  \Y(C \ \vec{v}) = 
\record{a_1 = M_1[\vec{v}/\vec{x}, O/\self], \ldots, a_k = M_k[\vec{v}/\vec{x}, O/\self]}
$
where $\Y \defEq \lambda f. (\lambda x f(xx))(\lambda x f(xx))$ is Curry's fixed point combinator and equality is conversion. Then sending a message $a$ to the object $O$ is defined  as record selection:
$\send{O}{a} \defEq O.a$. Indeed, for $i \in \Set{1,\ldots,k}$:
\[\send{O}{a_i} = (\record{a_1 = M_1[\vec{v}/\vec{x}, O/\self], \ldots, a_k = M_k[\vec{v}/\vec{x}, O/\self]}).a_i = M_i[\vec{v}/\vec{x}, O/\self].\]
For example let us consider the class:
$\PointClass \defEq \lambda x \ \lambda \self. \record{\textsf{X} = x, \ \get = \self.\textsf{X}}$,
such that $\textsf{X}$ is the  field holding the position of a one dimensional point (the object state), $\get$ returns the position of its point. 
Looking at the typings of 
$\PointClass (3) = \lambda \self. \record{\textsf{X} = 3, \ \get = \self.\textsf{X}}$, we have:
\[
\prooftree 
	\prooftree 
		\self:\omega \der 3 : \Int 
	\justifies
		\self:\omega \der \record{\textsf{X} = 3, \ \get = \self.\textsf{X}} : \record{\textsf{X}: \Int}
	\endprooftree
\justifies 
	\der \lambda \self. \record{\textsf{X} = 3, \ \get = \self.\textsf{X}} : \omega \arrow \record{\textsf{X}: \Int}
\endprooftree
\]
but also, setting $\sigma_1 \defEq \record{\textsf{X}: \Int}$:
\[
\prooftree 
	\prooftree 
		\self:\sigma_1 \der 3 : \Int \qquad  \self:\sigma_1 \der \self.\textsf{X}:\Int
	\justifies
		\self:\sigma_1 \der \record{\textsf{X} = 3, \ \get = \self.\textsf{X}} : \record{\textsf{X}: \Int, \ \get:\Int}
	\endprooftree
\justifies 
	\der \lambda \self. \record{\textsf{X} = 3, \ \get = \self.\textsf{X}} : \sigma_1 \arrow \record{\textsf{X}: \Int, \ \get:\Int}
\endprooftree
\]
so that, putting $\sigma_2 \defEq \record{\textsf{X}: \Int, \ \get:\Int}$, we have
\[\PointClass : \Int \arrow(\omega\arrow\sigma_1)  \ \inter \ (\sigma_1 \arrow \sigma_2)\]
and hence $\PointClass (3) : (\omega\arrow\sigma_1)  \ \inter \ (\sigma_1 \arrow \sigma_2)$.
It is folklore (see e.g. \cite{Dezani-Giovannetti-deLiguoro:Tokyo98}, Sec. 2) that with intersection types we can type 
$\Y \equiv \lambda f . (\lambda x.\; f(x\,x))(\lambda x.\; f(x\,x))$ 
by 
$(\omega\arrow\tau_1) \, \inter \, (\tau_1 \arrow \tau_2) \, \inter \cdots \, \inter \,(\tau_{n-1} \arrow \tau_{n}) \arrow \tau_{n}$
for any $n$ and $\tau_1,\ldots,\tau_{n}$; hence we conclude that
\[\pointObj \defEq \Y(\PointClass(3)) = \record{\textsf{X} = 3, \ \get = \pointObj.\textsf{X}}: \record{\textsf{X}: \Int, \ \get:\Int}\]
and therefore both $(\send{\pointObj}{\textsf{X}}): \Int$ and $(\send{\pointObj}{\get}): \Int$. Observe that neither typings 
of the class $\PointClass$ nor of the object $\pointObj$ are recursive types.

\newpage

Let $\vec{\const}$ be a vector of ground types and $\sigma_1,\ldots,\sigma_n$ be record types, i.e. the types of \emph{class instances}; then by generalizing from the above example we define:
\vspace{-4mm}
\[
 \mbox{\bf Non-recursive class type:} ~~~
 \kappa ::= \vec{\const}\arrow (\omega\arrow\sigma_1) \inter \bigcap_{i=1}^{n-1} (\sigma_i\arrow\sigma_{i+1}). 
\]
Note that $\kappa \leq  \vec{\const}\arrow (\sigma_i\arrow\sigma_{i+1})$ for any $i=1,\ldots,n-1$, but this typing would be not enough for typing objects obtained as fixed points of pre-instantiated classes.

\mypar {\bf Non-recursive mixins}
 are functions from non-recursive classes to non-recursive classes:
\[M \defEq \lambda \super \ \lambda \vec{x} \ \lambda \self. \ \Y (\super \ \vec{y}) \Override \record{a_1=M_1, \ldots , a_k=M_k}, ~~~
\vec{y} \subseteq \vec{x}.\]
Note that both $\self$ and $\super$ may occur free in the $M_i$. Let's consider the example of a mixin adding a second dimension to 
one dimension points of class $\PointClass$:
\[ \PointTwoDMx \defEq \lambda \super \ \lambda x \ \lambda y \ \lambda \self. \ 
	\Y (\super \ x) \Override \record{\textsf{Y} = y, \, \get = (c.\textsf{X}, \self.\textsf{Y})}
\]
The apparently equivalent term 
$\lambda \super \ \lambda x \ \lambda y \ \lambda \self. \ \Y (\super \ x) \Override 
\langle\textsf{Y} = y, \, \get = (x, \self.\textsf{Y})\rangle$ is coarser than $\PointTwoDMx$, because it ignores the usage of the state
variable $x$ by the superclass $\super$, which is a parameter. By simple computations we obtain:
\[\begin{array}{lll}
\PointTwoDMx ( \PointClass) & = & \lambda x \ \lambda y \ \lambda \self. \ 
	\Y (\PointClass \ x) \Override \record{\textsf{Y} = y, \, \get = ((\Y (\PointClass \ x)).\textsf{X}, \self.\textsf{Y})} \\ [1mm]
& = &  \lambda x \ \lambda y \ \lambda \self. \ 
	\record{\textsf{X} = x, \, \get = \ldots} \Override \record{\textsf{Y} = y, \, \get = (x, \self.\textsf{Y})} \\ [1mm]
& = & \lambda x \ \lambda y \ \lambda \self. \  
	\record{\textsf{X} = x, \, \textsf{Y} = y, \, \get = (x, \self.\textsf{Y})}
\end{array}\]
By adding product types (for exemplification purposes) we easily derive that $ \PointTwoDMx (\PointClass)$ has type:
\[ \Int \arrow \Int \arrow (\omega \arrow \record{\textsf{X} : \Int, \, \textsf{Y} : \Int}) \inter
(\record{\textsf{X} : \Int, \, \textsf{Y} : \Int} \arrow
\record{\textsf{X} : \Int, \, \textsf{Y} : \Int, \, \get : \Int \times \Int})\]
which is a non-recursive class type.
In general we conclude that a non-recursive mixin has type $\kappa_1\arrow\kappa_2$, where $\kappa_1,\kappa_2$ are non-recursive class types.

\if false
\noindent
The mixin $\PointTwoDMx$ adds a new field $\textsf{Y}$ and redefines the method $\get$. 
So it is not surprising that the meaning of
 $\get$  changes w.r.t. its meaning in the parent class $\PointClass$. But consider the mixin:
\[\FooMx \defEq  \lambda \super \ \lambda x \ \lambda \self. \ \Y (\super \ x) \Override \record{\textsf{X} = \trueVal}\]
Then we can form the class:
\[\begin{array}{lll}
\FooMx \ \PointClass & = &  \lambda x \ \lambda \self. \ \Y (\PointClass \ x) \Override \record{\textsf{X} = \trueVal} \\ [1mm]
& = & \lambda x \ \lambda \self. \record{\textsf{X} = x, \, \get = (\Y (\PointClass \ x)).\textsf{X} } \Override \record{\textsf{X} = \trueVal} \\ [1mm]
& = & \lambda x \ \lambda \self. \record{\textsf{X} = \trueVal, \, \get = (\Y (\PointClass \ x)).\textsf{X} }
\end{array}\]
Now let's define the object $\fooObj \defEq \Y (\FooMx \  \PointClass \ 3)$; then we have:
\[\fooObj = \record{\textsf{X} = \trueVal, \, \get =\pointObj.\textsf{X} } 
	= \record{\textsf{X} = \trueVal, \, \get = 3 }.
\]
The meaning of $\get$ is changed in the class
$\FooMx \ \PointClass$ w.r.t. its original meaning in $\PointClass$ because the overriding of the member $\textsf{X}$ doesn't affect the actual value of $\self$ on which $\get$ depends, that remains the definition of $\textsf{X}$ in $\PointClass \ x$. This is a consequence of the fact that recursive records model static binding of self, and not dynamic dispatch.

\noindent
Even if the class $\FooMx \ \PointClass$ and the object $\fooObj$ do not seem to make sense in a typed context, we can type them in our system, that  is a polymorphic type assignment system. 

Indeed, let $\sigma_3 \defEq \Int \arrow(\omega\arrow\sigma_1)  \ \inter \ (\sigma_1 \arrow \sigma_2)$ and
$\Gamma_1 = \Set{\super:\sigma_3, x:\Int, \self:\omega}$; then
\[
\prooftree
	\Gamma_1 \der \Y \ (\PointClass \ x): \record{\textsf{X}: \Int, \ \get:\Int} \qquad
	\Gamma_1 \der \record{\textsf{X} = \trueVal} : \record{\textsf{X} : \Bool}
\justifies
	\Gamma_1 \der \Y \ (\PointClass \ x) \Override \record{\textsf{X} = \trueVal}: \record{\textsf{X}: \Bool, \, \get : \Int }
\endprooftree
\]
and hence $\FooMx : \sigma_3 \arrow \Int \arrow \omega \arrow \record{\textsf{X}: \Bool, \, \get : \Int }$. Following
the same pattern of the typing derivation of the object $\pointObj$, we conclude that $\fooObj : \record{\textsf{X}: \Bool, \, \get : \Int }$, as expected.
\fi

\medskip\noindent {\bf Recursive classes and mixins.}
Recursive classes and mixins are necessary whenever the resulting class is used to generate a new object of the same class (see the example with movable points below). Recursive classes have an extra parameter $\myClass$:
\[ C' \defEq \lambda \myClass \ \lambda \vec{x} \ \lambda \self.   \record{a_1=M_1, \ldots , a_k=M_k}.\]
The fixed point of a recursive class is a class; so to instantiate a recursive class to an object, a double fixed point is needed:
$O' \defEq \Y ((\Y \, C') \vec{v})$. Because of this, types of recursive classes have the shape:
\[\mbox{\bf Recursive class type:} ~~~ \kappa'::=(\omega\arrow\kappa_1) \inter \bigcap_{i=1}^{n-1} (\kappa_i\arrow\kappa_{i+1})\]
where $\kappa_1,\ldots,\kappa_n$ are non-recursive class types. We observe that all $\kappa_i$ begin by the same the same $\vec{t}$ and
include record types $\sigma_{i,1},\ldots,\sigma_{i,n_i}$ that can be choosed so that $\sigma_{i,j} \geq \sigma_{i,j+1}$, to form a descending chain.

\noindent
The object $O'$ is a record, that can be merged in a mixin with a record 
$\Delta(\super)(\myClass)(\vec{x})(\self) \equiv \record{a_1=M_1, \ldots , a_k=M_k}$ of added/overridden members:
\[M' \defEq \lambda \super \ \lambda \myClass \ \lambda \vec{x} \ \lambda \self. \ \Y ((\Y \,\super) \ \vec{y}) \Override \Delta(\super)(\myClass)(\vec{x})(\self), ~~~
\vec{y} \subseteq \vec{x}.
\]
As an example, consider the following mixin:
\[\MovableMx \defEq \lambda \super \ \lambda \myClass \ \lambda x \ \lambda \self.   \ \Y ((\Y \,\super) \, x) \Override 
	\record{\textsf{move} = \lambda dx. \ \Y ((\Y \,\myClass) \, (\self.X + dx) )}
\]
Now let $\RecPointClass \defEq \lambda \myClass. \ \PointClass$ be a (vacuously) recursive version of $\PointClass$, and consider the object
$\movableObj \defEq \ \Y \, (\Y\,(\MovableMx \ \RecPointClass) \, 3)$. Then for example we have:
\[\begin{array}{lll}
& (\send{\movableObj}{\textsf{move}}) (4) \\ [1mm]
= &	((\Y (\Y \,\RecPointClass \ 3) \Override  \\ [1mm]
&	\hspace{1cm}  \record{\textsf{move} = \lambda dx. \ \Y (\Y \,(\MovableMx \ \RecPointClass) \, (\movableObj.X + dx))})
	\Leftarrow \textsf{move})
	(4) \\ [1mm]
= & (\record{\textsf{X} = 3, \, \get = \ldots, \,
	 \textsf{move} = \lambda dx. \ \Y (\Y \,(\MovableMx \ \RecPointClass) \, (\movableObj.X + dx))}.\textsf{move}) (4) \\ [1mm]
= &  \Y (\Y \,(\MovableMx \ \RecPointClass) \, (\movableObj.X + 4)) \\ [1mm]
= & \Y ((\Y \,(\MovableMx \ \RecPointClass)) \, (3 + 4)) \hspace{5cm} \mbox{since $~~\movableObj.X = 3$}\\ [1mm]
= & \record{\textsf{X} = 7, \, \get = \ldots, \,
	 \textsf{move} = \ldots}.
\end{array}
\]
To see a possible typing of $\MovableMx$, let $\kappa_1 \defEq \Int \arrow (\omega\arrow\sigma_1)  \ \inter \ (\sigma_1 \arrow \sigma_2)$, which we know to be a type of the class $ \PointClass$; then  $\omega\arrow \kappa_1$ is a type of $\RecPointClass$. Now
\[\prooftree
	\prooftree
		x:\Int \der \Y \  \RecPointClass  : \kappa_1 \qquad x:\Int \der x:\Int
	\justifies
		x:\Int \der \Y \  \RecPointClass  \ x: (\omega\arrow\sigma_1)  \ \inter \ (\sigma_1 \arrow \sigma_2)
	\endprooftree
\justifies
		x:\Int \der \Y (\Y \  \RecPointClass  \ x): \sigma_2 = \record{\textsf{X}:\Int,\, \get:\Int}
\endprooftree
\]
and therefore, with $\textsf{X}$ and $\get$ being different from $\textsf{move}$, and $\super$ being bound to $ \RecPointClass$ in
$\movableObj$ we have:
\[\super: \omega \arrow \kappa_1, \, x:\Int \der \Y (\Y \  \super  \ x) 
\Override \record{\textsf{move} = \lambda dx. \ \Y ((\Y \,\myClass) \, (\self.X + dx) }: \sigma_2.\]
Let $\Gamma_2 = \Set{\super: \omega\arrow\kappa_1, \, \myClass: \omega\arrow\kappa_1, \, x: \Int, \, \self: \sigma_1}$; then,
since we have:
\[ 
\prooftree
	\Gamma_2, dx:\Int \der \self.\textsf{X}: \Int \qquad \Gamma_2, dx:\Int \der dx:\Int
\justifies
	\Gamma_2, dx:\Int \der \self.\textsf{X} + dx: \Int
\endprooftree
\]
it follows that:
\[
\prooftree
\prooftree
	\prooftree
		\Gamma_2, dx:\Int \der \Y \ \myClass : \kappa_1 = \Int \arrow (\omega\arrow\sigma_1)  \ \inter \ (\sigma_1 \arrow \sigma_2)
	\justifies
		\Gamma_2, dx:\Int \der (\Y \ \myClass) \, (\self.X + dx) : (\omega\arrow\sigma_1)  \ \inter \ (\sigma_1 \arrow \sigma_2)
	\endprooftree
\justifies
	\Gamma_2, dx:\Int  \der \Y ((\Y \ \myClass) \, (\self.X + dx)): \sigma_2
\endprooftree
\justifies
	\Gamma_2, \der \lambda dx.\,\Y ((\Y \ \myClass) \, (\self.X + dx)): \Int \arrow \sigma_2
\endprooftree
\]
and we conclude that: 
\[\Gamma_2 \der \Y (\Y \  \super  \ x) 
\Override \record{\textsf{move} = \lambda dx. \ \Y ((\Y \,\myClass) \, (\self.X + dx) }: \sigma_2 \inter \record{\textsf{move}:\Int\arrow\sigma_2},
\]
so that by taking $\kappa_2 \defEq \Int \arrow (\sigma_2 \inter \record{\textsf{move}:\Int\arrow\sigma_2})$ we can derive the typing:
\[\MovableMx : (\omega\arrow\kappa_1) \arrow (\omega\arrow\kappa_1) \inter (\kappa_1\arrow \kappa_2),
\]
that is of the form $\kappa_1' \arrow \kappa_2'$ where $\kappa_1',\kappa_2'$ are recursive class types.

\newpage

\noindent{\bf Mixin composition.} Since mixins are just functions from classes to classes, the natural way to compose them is by functional composition. To make it easier to compare with definitions in \cite{BrachaC90}, we compute their composition up to conversion. For $i=1,2$ let 
\[M_i \defEq \lambda \super_i \ \lambda \self_i. \ \Y (\super_i) \Override \Delta_i(\super_i)(\self_i)\]
be a pair of (non-recursive) mixins, where $\Delta_i(\super_i)(\self_i)$ is the record of added/overridden methods, and we forget about the state for simplicity. Then, for a suitably
non-recursive class $C$ we have:
\[\begin{array}{lll}
(M_2 \circ M_1)(C) & = & M_2(M_1(C)) \\ [1mm]
& = & \lambda \self_2. \ \Y (M_1(C)) \Override \Delta_2(M_1(C)))(\self_2) \\ [1mm]
& = & \lambda \self_2. \ M_1(C)(\Y (M_1(C))) \Override \Delta_2(M_1(C)))(\self_2) \\ [1mm]
& = & \lambda \self_2. \ \Y(C) \Override \Delta_1(C)(\Y (M_1(C))) \Override \Delta_2(M_1(C)))(\self_2) \\ [1mm]
\end{array}\]
so that we obtain:
\[M_2 \circ M_1 = \lambda \super \ \lambda \self. \ \Y(\super) \Override \Delta_1(\super)(M_1(\super)) 
	\Override \Delta_2(M_1(\super))(\self),\]
which is essentially Bracha's wrapper composition in \S 5.2 of \cite{BrachaThesis}. In the slightly more complex case of recursive mixins (without the state):
\[M'_i \defEq \lambda \super_i \ \lambda \myClass_i \ \lambda \self_i. \ \Y (\Y \,\super_i) \Override 
	\Delta'_i(\super_i)(\myClass_i)(\self_i)
\] composition $M'_2 \circ M'_1$ is given by:

 \begin{tabbing} \hspace{3cm}
 $\lambda \super$ \=  $\lambda \myClass \ \lambda \self. \  \Y(\Y \,\super)$ \= $\Override$ \\
 \>  $\Delta'_1(\Y (M'_1 \super))$ \= $(\Y (\Y (M'_1 \super))) \Override \Delta'_2(M'_1 \super)(\myClass)(\self)$.
 \end{tabbing} In spite of their complex shape, recursive mixins have arrow types with recursive-record types as both left and right-hand sides. It follows that
 their composition into a {\em linearized} form \cite{BrachaC90} is just functional composition, that combined with the subtyping relation can be typed by: 
 \[
 \prooftree
 	\der M_1': \kappa_1' \arrow \kappa_2' \qquad \der M_2': \kappa_3' \arrow \kappa_4' \qquad  \kappa_2' \leq \kappa_3'
\justifies
	\der M_2'\circ M_1' : \kappa_1' \arrow \kappa_4' 
\endprooftree
\]

\mypar {\bf Subtype properties.}
In general a mixin abstraction
\begin{equation}\label{eq:mixins-sub}
 \lambda \super.\lambda \myClass. \lambda \vec{x}. \lambda \self. 
(\Y ((\Y \,\super) \, \vec{y} )) \Override \Delta(\super)(\myClass)(\vec{x})(\self),\qquad \vec{y}\subseteq\vec{x}
\end{equation}
has a type of the form $\kappa'_1 \to \kappa'_2$, and the instances of classes $\kappa'_i$ have record-type $\sigma_i$
for $i \in \{1, 2\}$, say, so that  $\sigma_2$ is the type of 
body of the abstraction (\ref{eq:mixins-sub}).
Inequation $\sigma_2 \leq \sigma_\Delta$
, where $\sigma_\Delta$ is the type of $\Delta(\super)(\myClass)(\vec{x})(\self)$, holds as a direct consequence of the rule $(\Override_r)$. However, $\sigma_2 \leq \sigma_1$ is only satisfied, if no label of the class $\super$ was overwritten. 
Therefore, our mixin encoding is more powerful  \cite{Ostermann02} than the existing approach of abstract base class inheritance. Analogously to Cook et al.~\cite{CookHC90}, we pinpoint the distinction between mixin application and subtyping.

\def\int{\textit{int}}
\def\Unit{\textit{Unit}}
\def\get{\text{get}}
\def\set{\text{set}}
\def\shift{\text{shift}}
\def\move{\text{move}}
\def\new{\text{new}}
\def\enquote#1{``#1''}

\newcommand{\Point}{\textsf{Point}}
\newcommand{\Movable}{\textsf{Movable}}
\newcommand{\SetAdapter}{\textsf{SetAdapter}}

\newcommand{\tauany}[2]{\ensuremath{\tau^{\text{#1}}_{\text{#2}}}}

\newcommand{\sigmainstance}[1]{\ensuremath{\sigma^{\text{#1}}_{\text{instance}}}}
\newcommand{\pipreinstance}[1]{\ensuremath{\pi^{\text{#1}}_{\text{pre-instance}}}}
\newcommand{\kappaconstructor}[1]{\ensuremath{\kappa^{\text{#1}}_{\text{constructor}}}}

\newcommand{\tauinst}[1]{\tauany{#1}{intstance}}

\newcommand{\tauci}{\tauinst{class}}

\newcommand{\taumi}{\tauinst{mixin}}
\newcommand{\taudi}{\tauinst{$\Delta$}}

\section{Mixin Features In Existing Programming Languages}
\label{sec:java}

In this section we implement mixins, presented in the lambda-calculus with records and a record-merge operator, using standard features of modern Object Oriented (OO) languages. The correspondence between calculus and programs is exemplified, in particular the correspondence between typability in lambda calculus and typability in Java like type systems.

Applying the theoretical model to a practically useful programming language, we face several challenges. First, mixins are not native concepts in mainstream OO languages. Code-generation and the regularity of a design pattern exploiting existing features, namely delegation, abstract classes, inheritance and generics, can compensate for this deficit. In Java-like languages inheritance alone is incapable to implement all mixin features due to restrictions of overriding (cf. Section \ref{sec:interpretations}) and overloading methods. Second, real languages use references and side effects which are not present in the presented calculus. To approximate this, state is modeled functionally in the calculus, while access to state is restricted only to $\textit{get}$ and $\textit{set}$ methods in the real language, e.g. Java. We consider the following subset of the presented calculus:

\begin{enumerate}
\item We consider recursive classes $C$ with an aggregated state $x$:\\
$C \defEq \lambda \myClass.\lambda x.\lambda \self. \record{a_1=M_1, \ldots , a_k=M_k}$
\item \label{restr:nostate} We consider recursive mixins $M$ that map recursive classes to recursive classes and do no add state: \\
$M \defEq \lambda \super.\lambda \myClass.\lambda x.\lambda \self. \ \Y ((\Y \,\super) \, x) \Override \record{a_1=M_1, \ldots , a_k=M_k}$
\item State, passed as variable $x$, is directly accessed only by $\textit{get}$ and $\textit{set}$ methods.
\item \label{restr:tuple} Each method that modifies the state of the underlying object returns a tuple $(\textit{newState}, \textit{result})$, where $\textit{newState}$ is the new state of the object (updated value to pass as x, not the updated object) after the method call and $\textit{result}$ is the result of the method call.
\item \label{restr:new} Each class and each mixin contains the special method $\textit{new} = \lambda x'.\Y (\myClass \, x')$
\end{enumerate}

In addition the above restrictions, we capture common subterms and decompose tuples using the \enquote{$\textsf{let ... in}$} construction, which only serves as syntactic sugar and disallows any recursive definition.

Restriction \ref{restr:nostate} is used to separate the concerns of storing data in classes and add/replace methods in mixins. Additionally, restrictions \ref{restr:nostate} and \ref{restr:tuple} are used for seamless delegation of methods. 
Restrictions \ref{restr:tuple} and \ref{restr:new} are used to implement functional state update. As an example, we model the invocation of method $a$ in an object $o$ that updates the state of $o$ and returns a result $r$ by 
$$
\textsf{let } (x, r) = o.a \textsf{ in } 
(\textsf{let } o' = o.new(x) \textsf{ in } \ldots)
$$
After the invocation of $a$, the variable $o'$ is bound to a new instance of the underlying class with an updated state. Note that the underlying class of $o$ is not expressed explicitly.

There are various more advanced approaches for modeling state in typed $\lambda$-calculi, e.g. reference cells \cite{DP2000} or Monads \cite{KiselyovLaemmel05}. The presented version using $\textit{get}$ and $\textit{set}$ methods, is chosen for simplicity and compatibility to the type system.

The remaining section describes the relationship between mixins in the presented calculus and well typed code implementing mixin features. This relationship is a necessary step for a future extension of manual mixin application to provably reliable automatic synthesis. In particular, the following correspondences are demonstrated:
\begin{itemize}
\item Recursive classes in the presented calculus and classes in OO languages.
\item $\Delta$-terms (records of updated methods) of recursive mixins and design pattern based $\Delta$-definitions.
\item Mixin application and delegation based composition of classes and mixin implementations.
\end{itemize}

\newpage

To illustrate those correspondences we use pseudo-code as an intermediate layer between calculus and real code. For brevity, implementation code is illustrated via UML diagrams. Example implementations in Java and C\# are available in online \footnote{\url{http://www-seal.cs.tu-dortmund.de/seal/downloads/research/ITRS2014PostProc.zip}}.

\begin{figure}
  \centering
  \begin{subfigure}[b]{0.65\textwidth}
    \flushleft
    \begin{lstlisting}[escapeinside=//]
Class Point
State
 x: int
Definitions
 get: Unit /$\to$/ int
 get() = { self.x }
 set: int /$\to$/ Unit
 set(x') = { self.x := x' }
 shift: Unit /$\to$/ Unit
 shift() = { self.set(self.get() + 1) }
    \end{lstlisting}
    \caption{Pseudo-code of class $\Point$}
    \label{lst:point}
  \end{subfigure}
  \begin{subfigure}[b]{0.34\textwidth}
    \centering
    \includegraphics[scale=0.8]{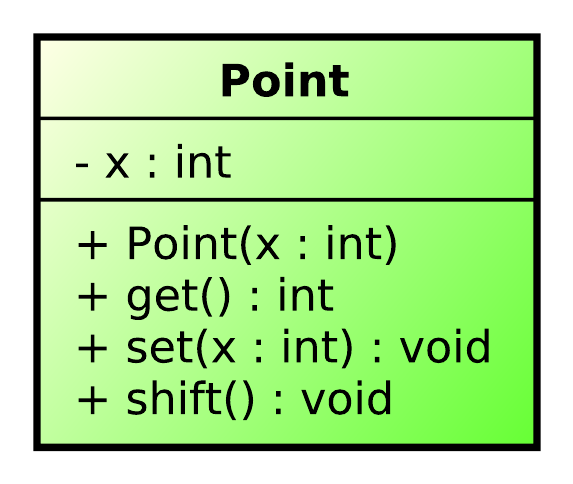}
    \caption{UML diagram of class $\Point$}
    \label{uml:point}
  \end{subfigure}
  \caption{Pseudo-code and UML diagram of class $\Point$}
  \label{fig:point_code}
\end{figure}

We extend the model-theoretic class $\PointClass$ from Section \ref{sec:intersection-types-lambda-rec} to the new class $\Point$ for interpreting the 
stateful Java/C\# class $\PointClass$, by adding
the methods $\textit{set}$ and $\textit{shift}$. The pseudo-code implementation as well as an UML representation of the Java/C\# implementation of $\Point$ are shown in Figure \ref{fig:point_code}. Respecting above restrictions to the presented calculus, we represent $\Point$ by the following term:
\begin{align*}
\Point = \lambda & \myClass.\lambda x.\lambda \self.\\
& \record{\get = x, \set = \lambda x'. (x', ()), \shift = \self.\set(\self.\get+1), \new = \lambda x'.\Y (\myClass \, x') }
\end{align*}

Note that interaction with the state $x$ of $\Point$ happens via the $\get$ and $\set$. We derive a type $\kappa_{\text{Point}}'$ out of the many types of $\Point$ using the following pattern:
\begin{align*}
\sigma_1^{\text{\Point}} =&\record{\get : \int, \set : \int \to (\int \times \Unit), \shift: \omega, \new: \int \to \omega}\\
\sigma_2^{\text{\Point}} =&\record{\get : \int, \set : \int \to (\int \times \Unit), \shift: (\int \times \Unit), \new: \int \to \omega}\\
\sigma_3^{\text{\Point}} =& \record{\get : \int, \set : \int \to (\int \times \Unit), \shift: (\int \times \Unit), \new: \int \to \sigma_2^{\text{\Point}}}\\
\kappa_1^{\text{\Point}} =& \int \to (\omega\to\sigma_1^{\text{\Point}})\inter(\sigma_1^{\text{\Point}}\to\sigma_2^{\text{\Point}})\\
\kappa_2^{\text{\Point}} =& \int \to (\omega\to\sigma_1^{\text{\Point}})\inter(\sigma_1^{\text{\Point}} \to \sigma_2^{\text{\Point}}) \inter (\sigma_2^{\text{\Point}} \to\sigma_3^{\text{\Point}})\\
\kappa_{\text{Point}}' = &(\omega\to\kappa_1^{\text{\Point}})\inter(\kappa_1^{\text{\Point}} \to \kappa_2^{\text{\Point}})  
\end{align*}
\noindent The type $\sigma_3^{\text{\Point}}$ describes an instance of $\Point$ after the double fixed point construction
fixing $\myClass$ and $\self$. The type  $\sigma_2^{\text{\Point}}$ describes the intermediate term resulting from the first fixed point construction that fixes $\myClass$.

\begin{figure}[ht]
  \centering
  \begin{subfigure}[b]{0.35\textwidth}
    \flushleft
    \begin{lstlisting}[escapeinside=//]
Mixin Movable(C)
Requirements
 get: Unit /$\to$/ int
 set: int /$\to$/ Unit
Definitions
 move: int /$\to$/ Unit
 move(dx) = 
  { super.set(super.get()+dx) }
    \end{lstlisting}
    \caption{Pseudo-code of mixin \Movable}
    \label{lst:movable}
  \end{subfigure}
  \begin{subfigure}[b]{0.639\textwidth}
    \centering
    \includegraphics[scale=0.49]{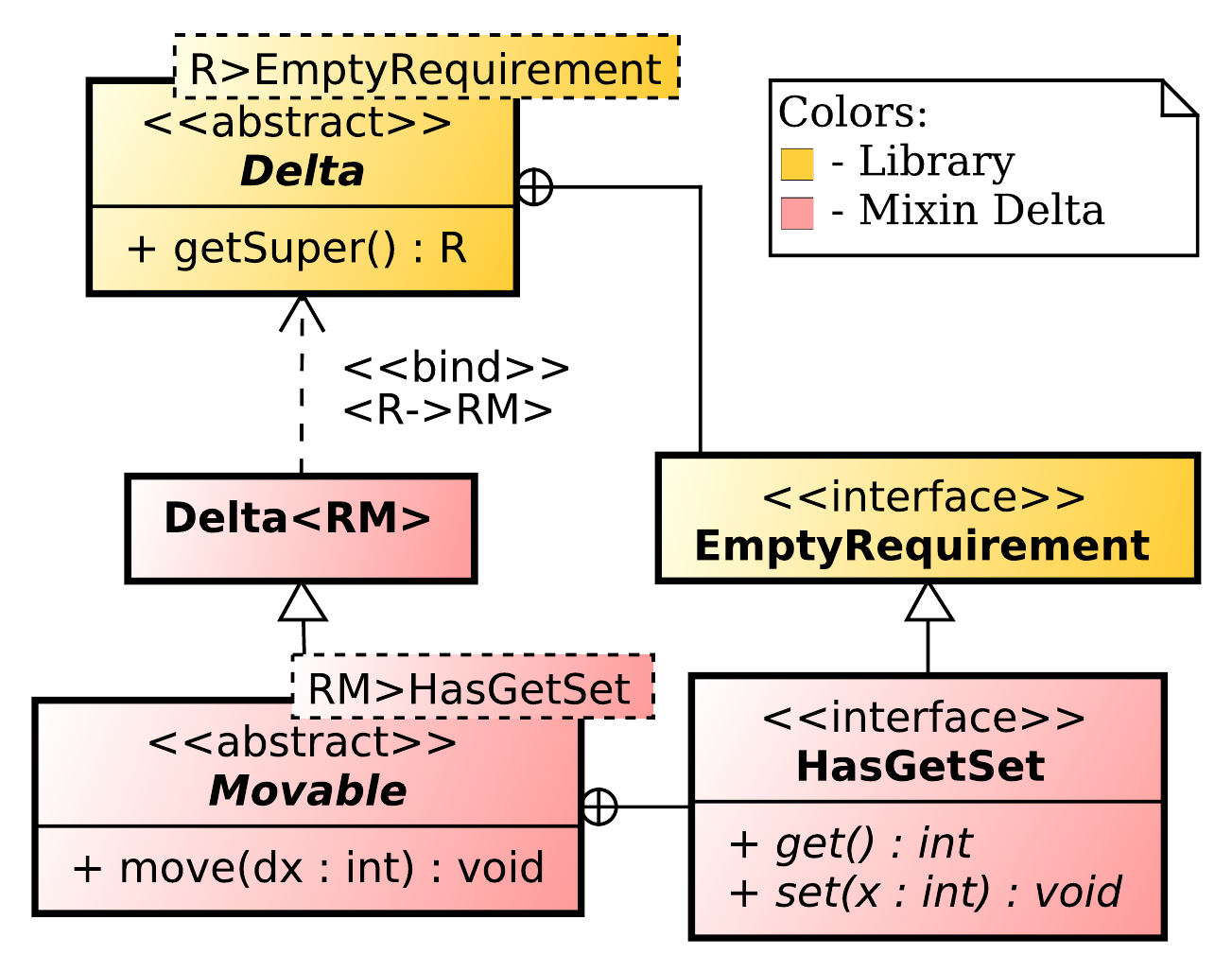}
    \caption{UML Diagram of mixin \Movable}
    \label{uml:movable}
  \end{subfigure}
  
  \begin{subfigure}[b]{0.35\textwidth}
    \flushleft
    \begin{lstlisting}[escapeinside=//]
Mixin SetAdapter(C)
Requirements
 set: int /$\to$/ Unit
Definitions
 set: Point /$\to$/ Unit
 set(p) = 
  { super.set(p.get()) }
    \end{lstlisting}
    \caption{Pseudo-code of mixin \SetAdapter}
    \label{lst:setadapter}
  \end{subfigure}
  \begin{subfigure}[b]{0.64\textwidth}
    \centering
    \includegraphics[scale=0.5]{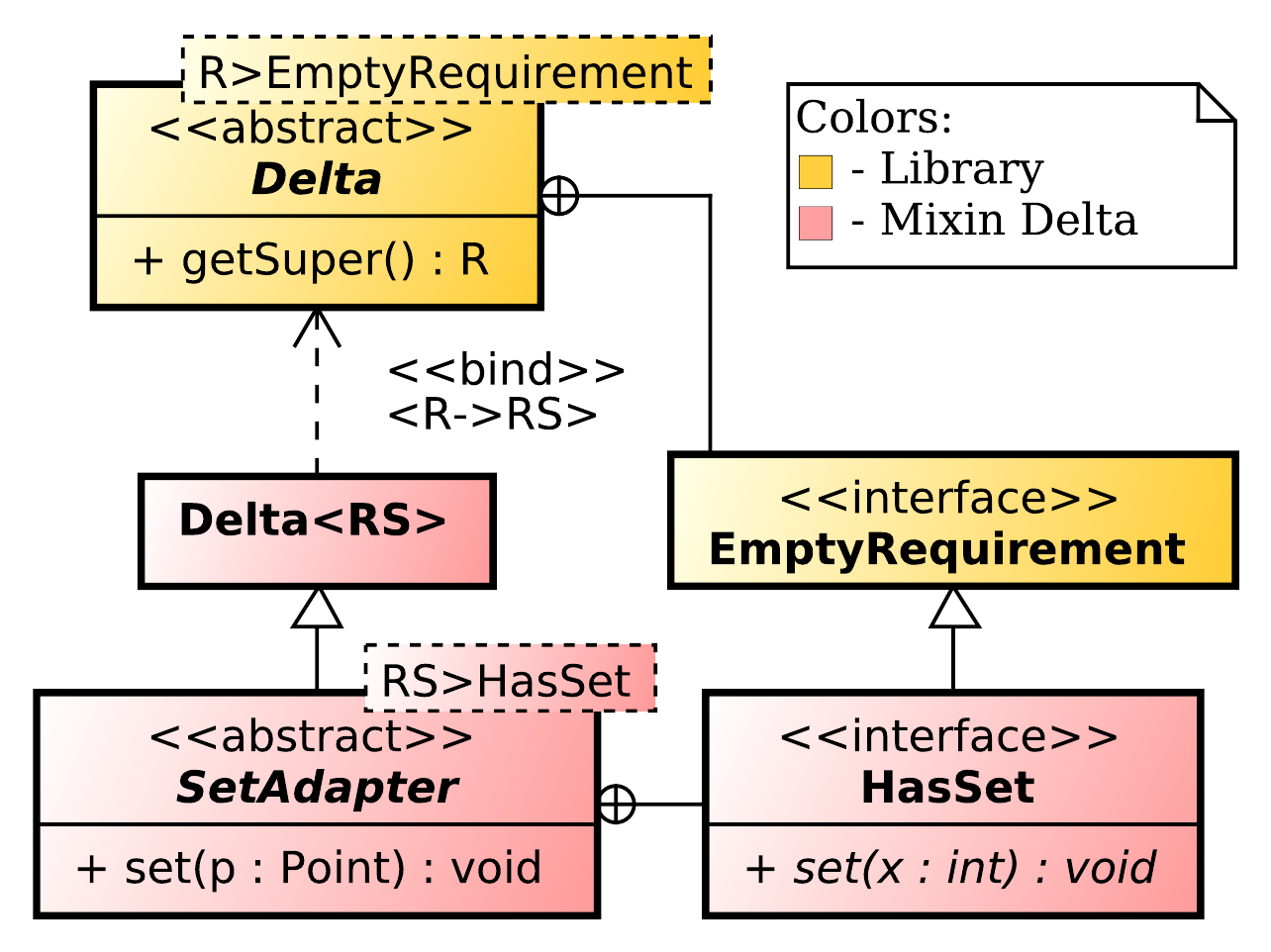}
    \caption{UML Diagram of mixin \SetAdapter}
  \end{subfigure}
  \caption{Pseudo-code and UML diagrams of mixins $\Movable$ and $\SetAdapter$}
  \label{fig:mixins_code}
\end{figure}

\noindent

Next, we define two mixins $\Movable$ and $\SetAdapter$ presented in Figure \ref{fig:mixins_code}. $\Movable$ requires the $\get$ and $\set$ methods to define the $\move$ method, moving the current point by $dx$. \textit{SetAdapter} redefines the $\set$ method to handle an instance of $\Point$ as input. Note that $\super$ is used to access methods of the corresponding mixin argument.

\noindent Respecting above restrictions to the presented calculus, we represent $\Movable$ by the following term:
\begin{align*}
\text{Movable} =  \lambda & \super.\lambda \myClass.\lambda x.\lambda \self.\\
& \textsf{let } c = \ \Y ((\Y \,\super) \, x) \textsf{ in } \\
& \textsf{let } \Delta = \record{move = \lambda dx. \ c.set(c.get + dx), \textit{new} = \lambda x'.\Y (\myClass \, x')} \textsf{ in } \\
& c \Override \Delta
\end{align*}
We derive a type $\kappa_{\text{\Point}}' \to \kappa_{\Movable}'$ out of the many types of $\Movable$ using the following pattern:
\begin{align*}
\sigma_1^{\text{\Movable}} = & \record{\get : \int, \set : \int \to (\int \times \Unit), \shift:  (\int \times \Unit)} \\
& \cap \record{\move : \int \to (\int \times \Unit),\new: \int \to \omega}\\
\sigma_2^{\text{\Movable}} = & \record{\get : \int, \set : \int \to (\int \times \Unit), \shift:  (\int \times \Unit)} \\
& \cap \underbrace{\record{\move : \int \to (\int \times \Unit),\new: \int \to \sigma_1^{\text{\Movable}}}}_{\text{type } \sigma_{\Delta}^{\text{\Movable}} \text{ induced by the mixin description}}\\
\kappa_1^{\text{\Movable}} = & \int \to (\omega \to \sigma_1^{\text{\Movable}})\\
\kappa_2^{\text{\Movable}} = & \int \to (\omega \to \sigma_1^{\text{\Movable}}) \cap (\sigma_1^{\text{\Movable}} \to \sigma_2^{\text{\Movable}}) \\
\kappa_{\text{\Movable}}' = & (\omega \to \kappa_1^{\text{\Movable}}) \cap (\kappa_1^{\text{\Movable}} \to \kappa_2^{\text{\Movable}})
\end{align*}

Note that the type of $\move : \int \to (\int \times \Unit)$ follows a uniform type structure not bound to the mixin $\Movable$ and, therefore, allows for seamless delegation.

Similarly, $\SetAdapter$ is represented by the following term and is typed by $\kappa_{\text{\Movable}}' \to \kappa_{\text{SetAdapter}}'$:
\begin{align*}
\text{SetAdapter} = \lambda & \super.\lambda \myClass.\lambda x.\lambda \self. \\
& \textsf{let } c = \ \Y ((\Y \,\super) \, x) \textsf{ in } \\
& \textsf{let } \Delta = \record{set = \lambda p.c.set(p.get), \textit{new} = \lambda x'.\Y (\myClass \, x')} \textsf{ in } \\
& c \Override \Delta
\end{align*}

\ifx true false
\begin{align*}
\sigmainstance{$\Delta$-SetAdapter} = & \; \sigmainstance{Point} \inter (\sigma_2\to\sigma_6)~~~~~~~~~\mbox{where}\\
\sigma_6 =& \; \record{\get : \int, \set :\sigma_2 \to (\int \times \Unit), \shift: (\int \times \Unit), \move : \int \to (\int \times \Unit),\new: \int \to\omega}\\
\sigmainstance{SetAdapter} = & \sigmainstance{Point}\inter \sigmainstance{$\Delta$-SetAdapter}\\
\kappa_5 =& \; \int\to \sigmainstance{SetAdapter}\\
\sigma_7 =& \;  \record{\get : \int, \set :\sigma_2 \to (\int \times \Unit), \shift: (\int \times \Unit), \move : \int \to (\int \times \Unit),\new: \int \to\kappa_5}\\
\kappa_6 =& \; \int\to \sigmainstance{SetAdapter} \\
\kappa'^{\text{SetAdapter}} =& \;  \kappa'^{\text{Point}} \inter (\kappa_2\to\kappa_5)\inter(\kappa_5\to\kappa_6) \\
\end{align*} 
\fi

\begin{align*}
\sigma_1^{\text{\SetAdapter}} =& \record{\get : \int, \shift:  (\int \times \Unit), \move : \int \to (\int \times \Unit)} \\
& \cap \record{\set : \record{\get : \int} \to (\int \times \Unit), \new: \int \to \omega}\\
\sigma_2^{\text{\SetAdapter}} =& \record{\get : \int, \shift:  (\int \times \Unit), \move : \int \to (\int \times \Unit)} \\
& \cap \record{\set : \record{\get : \int} \to (\int \times \Unit), \new: \int \to \sigma_1^{\text{\SetAdapter}}}\\
\sigma_3^{\text{\SetAdapter}} =& \record{\get : \int, \shift:  (\int \times \Unit), \move : \int \to (\int \times \Unit)} \\
& \cap \underbrace{\record{\set : \record{\get : \int} \to (\int \times \Unit), \new: \int \to \sigma_2^{\text{\SetAdapter}}}}_{\text{type } \sigma_{\Delta}^{\text{\SetAdapter}} \text{ induced by the mixin description}}\\
\kappa_1^{\text{\SetAdapter}} = & \int \to (\omega \to \sigma_1^{\text{\SetAdapter}})\\
\kappa_2^{\text{\SetAdapter}} = & \int \to (\omega \to \sigma_1^{\text{\SetAdapter}}) \cap (\sigma_1^{\text{\SetAdapter}} \to \sigma_2^{\text{\SetAdapter}})\\
\kappa_3^{\text{\SetAdapter}} = & \int \to (\omega \to \sigma_1^{\text{\SetAdapter}}) \cap (\sigma_1^{\text{\SetAdapter}} \to \sigma_2^{\text{\SetAdapter}}) \cap (\sigma_2^{\text{\SetAdapter}} \to \sigma_3^{\text{\SetAdapter}})\\
\kappa_{\text{\SetAdapter}}' = & (\omega \to \kappa_1^{\text{\SetAdapter}}) \cap (\kappa_1^{\text{\SetAdapter}} \to \kappa_2^{\text{\SetAdapter}}) \cap (\kappa_2^{\text{\SetAdapter}} \to \kappa_3^{\text{\SetAdapter}})
\end{align*}

We encode $\Delta$-terms in the implementation. Each $\Delta$-term is represented by an abstract class extending the class \texttt{Delta}, which is available in an utility library. $\Delta$-term representations are parameterized over a requirement interface type. This type is used for the library method \texttt{getSuper}, which provides the instance with which the $\Delta$ implementation will be merged. This instance can be guaranteed to be at least of the required interface type. Required interface types may be declared for each $\Delta$ (e.g. \texttt{HasGetSet}). They are interfaces which, for reasons of structural regularity, extend the \texttt{EmptyRequirement} interface, also available in the utility library. $\Delta$-term representations remain parametric over the type of the term they are merged with in order to allow \texttt{getSuper} to return an instance of that type, rather than just an instance of the requirement interface type. This is necessary to reuse \texttt{getSuper} for delegation of methods propagated after mixin application, without mentioning them in the requirement interface.

\begin{figure}[h]
  \includegraphics[scale=0.45]{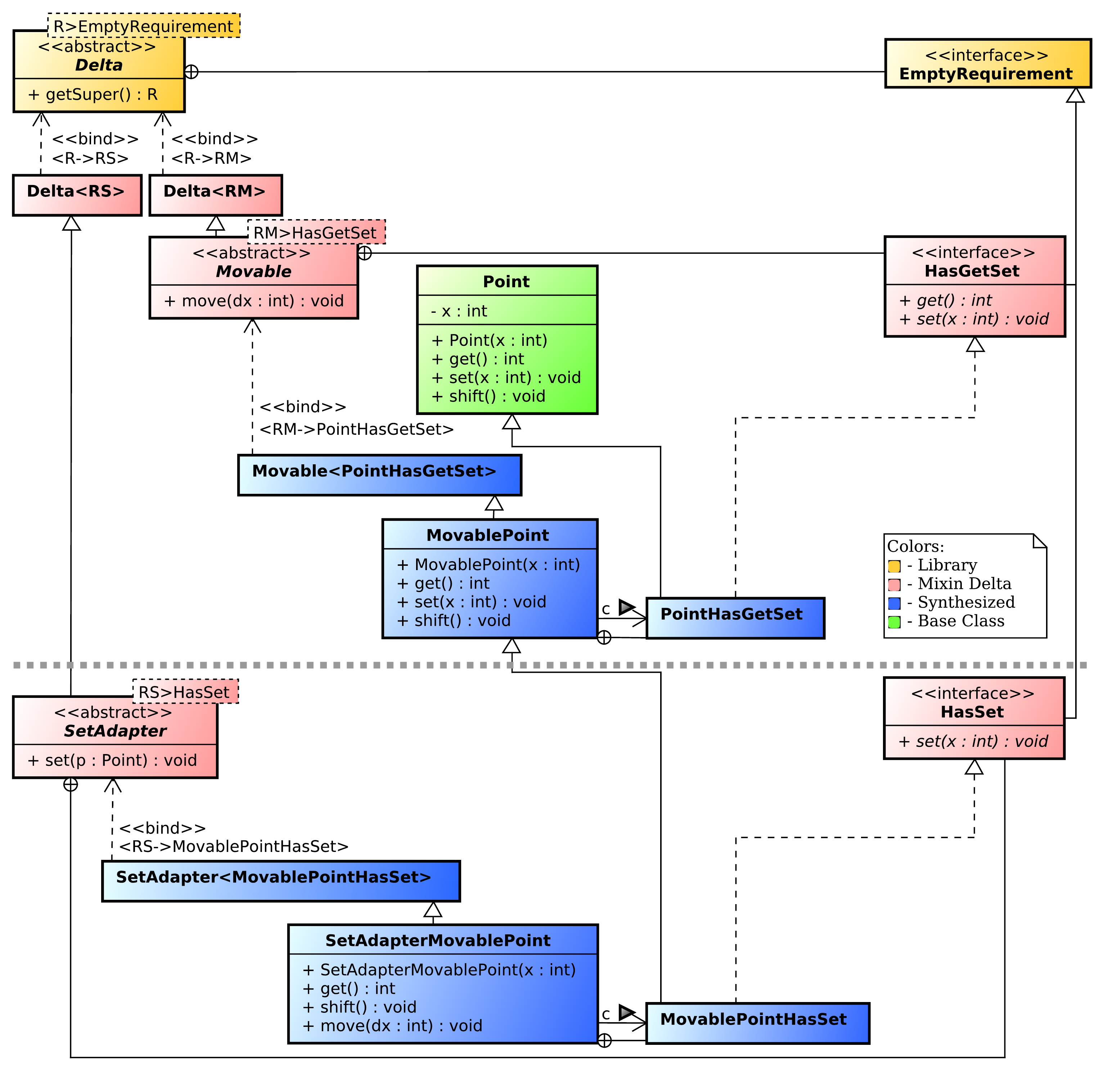}
  \caption{UML Diagram for synthesized code. The dotted line separates Movable(Point) from \\ SetAdapter(Movable(Point))}
  \label{uml:synthesized}
\end{figure}

Figure \ref{uml:synthesized} shows the implemented result of applying \texttt{Movable} to \texttt{Point} and (below the dashed line) the application of \texttt{SetAdapter} to the previous result. First, \texttt{MovablePoint} is obtained by creating a new class that inherits from \texttt{Movable}. It thereby satisfies $ \sigma_{n}^{\text{Mixin}} \leq \sigma_{\Delta}^{\text{Mixin}}$. Internally, \texttt{MovablePoint} delegates to the methods of a \texttt{PointHasGetSet} instance provided by the \texttt{getSuper} method inherited from \texttt{Delta}. Class \texttt{PointHasGetSet} extends \texttt{Point} and implements the requirement interface. The introduction of a new class is necessary in languages not supporting structural subtyping. Structurally $\texttt{Point} \leq \texttt{HasGetSet}$ does hold, but in a nominal type system \cite{CookHC90} subtype relationships have to be stated explicitly. The class \texttt{MovablePoint} propagates the non overwritten methods of \texttt{Point}, e.g. \texttt{shift}. Note that \texttt{shift} is not part of the requirement interface, but available since the type parameter of \texttt{Movable} is bound to \texttt{PointHasGetSet}.
Similarly, \texttt{SetAdapterMovablePoint} extends \texttt{SetAdapter} and delegates to \texttt{MovablePointHasSet}, which connects \texttt{HasSet} and \texttt{MovablePoint}.

The UML diagram reveals the very regular nature of mixin application, which in a practical setting can be automated, s.t. blue classes in the diagram are synthesized by a code generator and do not have to be implemented by hand. Code generation corresponds to $\Lambda_R$ reduction, performing the merge operation $\Override$ on the $\textsf{super}$ argument. In the case of Java and C\# it is to be performed outside of the implementation language, because the creation of new named classes is not a native language feature. 

\begin{figure}
 \centering
 \begin{subfigure}[t]{0.57\textwidth}
    \flushleft
    \begin{lstlisting}[escapeinside=//,numbers=left,numberstyle=\tiny]

p1 := new SetAdapter(Movable(Point))(1)
p2 := new Point(2)
p1.set(p2) 

p1.move(1)

p1.get()
    \end{lstlisting}
    \caption{Implementation pseudo-code}
    \label{lst:pseudo_usage}
  \end{subfigure}
  \begin{subfigure}[t]{0.419\textwidth}
    \begin{lstlisting}[escapeinside=//]
/$\textsf{let } C = (SetAdapter \circ Movable)(Point) \textsf{ in}$/
/$\textsf{let } p1 = \Y ((\Y \, C) 1) \textsf{ in}$/
/$\textsf{let } p2 = \Y ((\Y \, Point) 2) \textsf{ in}$/
/$\textsf{let } (x, r) = p1.set(p2) \textsf{ in}$/
/$\textsf{let } p1' = p1.new(x) \textsf{ in}$/
/$\textsf{let } (x', r') = p1'.move(1) \textsf{ in}$/
/$\textsf{let } p1'' = p1'.new(x') \textsf{ in}$/
/$p1''.get()$/
    \end{lstlisting}
    \caption{$\Lambda_R$-calculus}
    \label{lst:calculus_usage}
  \end{subfigure}
  \caption{Instantiation and usage of Mixins}
\end{figure}

We compare the usage of the pseudo-code implementation and calculus description in the example shown in Figures \ref{lst:pseudo_usage} and \ref{lst:calculus_usage}. Implementations in Java and C\# are also available online and almost identical to the pseudo-code implementation. The main differences between the pseudo-code and the $\Lambda_R$ version are the initial class instantiation and the different styles of passing state. In OO languages class instantiation is performed by the built-in operator \texttt{new}. In $\Lambda_R$ we model \texttt{new} using the fixpoint operator \textbf{Y} respectively (cf. Section \ref{sec:interpretations}). Passing state functionally requires explicit instantiation of copies of $p1$ with an updated state (lines 5 and 7). Observe that in $p1.set$ (line 4) is the new set method, however, $p1.move$ (line 6) uses the old set method, which is the required behavior. Therefore, the final result produced by both implementations (line 8) is $3 : \int$. Observing that $\sigma_2^{\text{\SetAdapter}}$ are $\sigma_3^{\text{\SetAdapter}}$ necessary for the twofold state update of $p1$, we derive $\der p1 : \sigma_3^{\text{\SetAdapter}}$ using the following derivation for $(\SetAdapter \circ \Movable)(\Point)$:
\[\prooftree 
	\prooftree 
	\der \Movable: \kappa_{\Point}' \to \kappa_{\Movable}' \qquad
	\der \SetAdapter: \kappa_{\Movable}' \to \kappa_{\SetAdapter}'
	\justifies \der \SetAdapter \circ \Movable: \kappa_{\Point}' \arrow \kappa_{\SetAdapter}'
	\endprooftree
	\qquad \der\Point:  \kappa_{\Point}'
\justifies \der (\SetAdapter \circ \Movable)(\Point): \kappa_{\SetAdapter}'
\endprooftree
\]

\newpage

\section{Conclusions and further work}

%
%

In this paper we provide a lambda-calculus with records and a record-merge operator to describe classes, mixins and mixin-applications in object-oriented languages featuring an early binding of method calls. We demonstrate its potential by composing well typed programs in Java and C\# that itself do not inherently support mixins. The delegation based approach in combination with subtyping turns out to be more flexible as well as powerful than inheritance and overloading. In future, we aim to develop a theory for automatically synthesizing code for mixin-applications using an approach introduced in \cite{BessaiDDM14}.

\bibliographystyle{eptcs}

\newpage

\bibliography{SemMixBiblio,SyntBibliography}

\end{document}